\documentstyle[aps,prl,preprint]{revtex}
\begin{document}
\draft

\title{Independent-Electron Model for the Phase
       of the Transmission Amplitude in Quantum Dots}
\author{G.\ Hackenbroich and H.\ A.\ Weidenm\"uller}
\address{Max-Planck-Institut f\"ur Kernphysik, 69117 Heidelberg, Germany}
\date{\today}
\maketitle

\begin{abstract}
Motivated by a recent experiment by Yacoby et al.\ [preprint, 1994], we
calculate magnitude and phase $\alpha$ of the transmission amplitude
through a quantum dot. We work in the Coulomb blockade regime, assume
the electrons not to interact, and consider tunneling through isolated
resonances. Assuming a non--resonant background, we find that $\alpha$
increases by $2 \pi$ over each resonance, with a sharp rise by $\pi$
over an energy interval much smaller than the thermal width. This is
consistent with the experimental data. Our assumptions can be tested by
further experiments.
\end{abstract}

\narrowtext
A direct measurement of the phase of the amplitude for transmission through
a quantum dot has recently been reported for the first time \cite{Ya94}. The
Coulomb repulsion of the electrons in the dot causes a sequence of isolated
resonances to appear in the transmission coefficient as the voltage on the
dot is changed. Each resonance corresponds to a different number of electrons
on the dot. In the experiment \cite{Ya94}, the phase $\alpha$ of the
transmission amplitude was measured in several of these resonances.
This was done using an Aharanov--Bohm (AB) type interference experiment
with a quantum dot embedded
in the left of the two arms of an AB ring. The ring was threaded by a magnetic
flux $\Phi$, and the flux dependence of the transmission through the ring
was measured. It was found (i) that the energy dependence of the phase
$\alpha$ is the same in all resonances studied, and that (ii) in each
resonance, the phase $\alpha$ increases {\it sharply} by ${\approx \pi}$.
The increase occurs over an energy interval much smaller than the width
of the resonance. Both observations seemed incompatible with the model
of non--interacting electrons \cite{Ya94}.

It is the purpose of this paper to present a theoretical framework for the
analysis of this and similar future experiments, and to analyse the data of
Ref. \cite{Ya94} in this framework. To describe the transmission
through the two--arm AB  device, we assume the electrons to move
independently. As a formal framework, we use the Landauer--B\"uttiker
approach \cite{Lan}. We allow for the existence of several channels
in either arm.
Discussing the results of Ref. \cite{Ya94} in this framework, we can
(i) easily explain the first observation. We also find (ii) an increase by
${\pi}$ of the phase over an energy interval which is small compared to
the width of the resonance. The steepness of the
increase depends on the values of the resonance and background parameters.
With a suitable choice for these values, we can reproduce the observed
behavior. Our interpretation can be tested by further experiments.

The periodic conductance oscillations of quantum dots coupled to external
leads by tunnel barriers have been explained in terms of the Coulomb blockade
model \cite{AL}. With $C$ the mutual capacitance between the dot and the rest
of the system, addition of a single electron to the dot requires a charging
energy $e^2/C$. At temperatures $T$ so small that $e^2/C \gg k_B T$, and for
sufficiently small driving voltages, electrons cannot enter the dot, and the
conductance is suppressed. This Coulomb blockade can be removed by applying
a plunger voltage to the dot. The charging energy then varies periodically with
this voltage, leading to periodic peaks in the conductance. For sufficiently
small dots made of semiconductors, this classical picture needs to be modified
because of the discreteness of the energy spectrum within the dot \cite{B}.
For the dot used in Ref. \cite{Ya94}, the average level spacing $\Delta E$ is
roughly similar to $e^2/C$, and $\Delta E$ must be taken into account in the
estimate for the period of the conductance peaks. The fact that
$\Delta E \gg k_B T$ suggests that the
predominant conduction mechanism is resonant tunneling through a single
electron level in the quantum dot. Therefore, we refer to the conductance
peaks as to resonances. The typical intrinsic width $\Gamma$ of each
resonance is much smaller than both the temperature $k_BT$ and the
average level spacing $\Delta E$.

In the actual experiment, the current through the AB ring was measured for
a fixed flux $\Phi$ as a function of the plunger voltage. On top of a large
background due to transmission through the right arm, a series of
well--separated resonances was observed. At several resonances, and for
several values of the plunger voltage in each resonance, the magnetic
flux was varied. The conductance was found to have a coherent component
oscillating periodically with $\Phi / \Phi_0$ where $\Phi_0 = h/e$ is
the elementary flux quantum. The change of phase of these oscillations as
the plunger voltage passes through a resonance defines the energy dependence
of the phase $\alpha$ of the transmission amplitude. It was found that
$\alpha$ changes by multiples of $2 \pi$ between subsequent resonances,
and that at each resonance, $\alpha$ increases sharply by ${\pi}$ on a scale
which is small compared to the thermal width of the resonance.

To model this experiment, we consider an AB ring coupled to two ideally
conducting leads. A quantum
dot embedded in the left arm is separated from the ring by two tunnel
barriers. The electrons are assumed not to interact. Both the ring and
the dot are allowed to contain some disorder. In linear response theory,
the dimensionless conductance $g = (h/e^2)G$ of the ring can be calculated
from the multi--channel Landauer formula \cite{Lan}
\begin{equation}
  g=2\int dE \left(-{\partial f \over \partial E}\right) \sum_{a,b=1}^N |t_{ab}
    (E)|^2 .
\label{E1}
\end{equation}
Here, $t_{ab}(E)$ is the transmission amplitude through the ring, taken at
energy $E$, for an electron entering the ring via channel $a$ in one lead,
and leaving it via channel $b$ in the other lead. The total number of
channels in either lead is denoted by $N$. The derivative of the Fermi
function $f$ is given by $- \big( \partial f/\partial E \big)
= (4 k_B T)^{-1} \cosh^{-2}((E - E_F)/2 k_B T)$, and $E_F$ is the Fermi energy
in the leads. A factor two accounts for the spin degeneracy of the electron.

The transmission amplitude $t_{ab}$ through the ring is the coherent sum
of the transmission amplitudes $t_{ab}^L$ and $t_{ab}^R$ through the left
arm and the right arm, respectively, $t_{ab} = t_{ab}^L + \exp ( 2i \pi \Phi
/ \Phi_0 ) t_{ab}^R$. We model the quantum dot as having one single--particle
resonance at energy $E_0$ and write $t_{ab}^L$ in the form
\begin{equation}
  t_{ab}^L = t_{ab}^{L0} -i t_a^L {\Gamma \over E- E_0+i \Gamma / 2}t_b^L ,
\label{E2}
\end{equation}
where $\Gamma$ denotes the intrinsic width of the resonance, and where
$t_{ab}^{L0}$ accounts for non--resonant transmission through
the dot outside the domain of the single--particle resonances.
(Possible mechanisms leading to non-zero
$t_{ab}^{L0}$ are for example a small ballistic contribution due to incomplete
pinching of the quatum dot or the accumulated tunneling transmission due to the
presence of far away resonances). The form of the resonance term follows
trivially from the Breit--Wigner multi--channel single--resonance
amplitude, and from the assumption that further amplitudes are needed to
describe the transmission of electrons from channel
$a$ to the dot, and from the dot to channel $b$. The complex amplitudes
$t_{ab}^R, t_{ab}^{L0}, t_a^L$ and $t_b^L$ vary slowly with energy and
are taken to be constant over the observed width of the resonance.
Inserting the form (\ref{E2}) of $t_{ab}$ into the Landauer formula
(\ref{E1}), and omitting
the background terms which are both energy and flux independent, we find
for the remaining contribution $g^0$ to the conductance,
\begin{eqnarray}
  g^0 & = & 2 \int dE \left(-{\partial f \over \partial E}\right)
  \sum_{a,b=1}^N
  \left\{ |t_a^L|^2 |t_b^L|^2 {\Gamma^2 \over (E-E_0)^2 + \Gamma^2/4}+
  \left(i t_{ab}^{L0} t_a^{L*}  t_b^{L*} {\Gamma \over E-E_0 -i \Gamma/2}
  +c.c.\right)
  \right. \nonumber \\
  & & \left.+\left( t_{ab}^{L0*}t_{ab}^R \exp(2 i \pi {\Phi \over \Phi_0} ) +
  c.c. \right)+
  \left(i t_{ab}^{R} t_a^{L*} t_b^{L*} {\Gamma \over E-E_0 -i \Gamma/2}
  \exp(2 i\pi {\Phi \over \Phi_0}) +c.c.\right) \right\}.
\label{E3}
\end{eqnarray}
Of the four terms on the right--hand side of Eq.\ (\ref{E3}), the first one
is the absolute square of the resonance term and has the usual Breit--Wigner
form. The three other terms describe the {\it interference} between pairs of
the following amplitudes: The resonance amplitude, the amplitude $t_{ab}^{L0}$
for non--resonant transmission through the left arm, and the amplitude
$t_{ab}^R$ for transmission through the right arm. Only the third and
fourth terms depend on the applied magnetic flux and are responsible for AB
oscillations. We note that the two interference terms which contain the
resonance amplitude dacay like
$1/|E- E_0|$ and dominate the Breit--Wigner resonance (the first term in
Eq.\ (\ref{E3})) for $|E - E_0| \gg \Gamma$. This
broadening of the resonance due to interference  will be seen to persist after
temperature--averaging.

To estimate the relative weight of the different contributions to Eq.\
(\ref{E3}) in the experiment, we note that the resistance of the left
arm containing the dot exceeds that of the right arm by a large factor.
This statement only reflects the fact that even at a resonance, the
electrons are required to tunnel through the dot. Outside the resonance,
transmission through the left arm is further inhibited. Therefore,
we expect the ratio
$|t_{ab}^R/t_{ab}^{L0}|$ to be significantly larger than
$|t_{ab}^R/t_a^Lt_b^L|$. Moreover, the first term in Eq. (\ref{E3})
is the square of a tunnel amplitude and should therefore be suppressed
in comparison with the fourth term. On the other hand, all
interference contributions are strongly supressed by thermal averaging
and/or by inelastic scattering and by the summation over the $N$ channels.
(This is why for the bare ring without the dot, the oscillating part of the
resistance comprises only about $10$ per cent of the average  resistance).
Taking account of all these arguments, we expect the first and the fourth
terms
in  Eq.\ (\ref{E3}) to be roughly of the same order. Both should be much
larger than the third term which in turn should dominate the second term.
This second term is, therefore, neglected. We do keep, however, the term
containing the interference with the non--resonant transmission amplitude
$t_{ab}^{L0}$. This term will be seen to determine the steepness of the
slope of the measured phase $\alpha$. After summing over channels,
$g^0$ can be written in the form
\begin{eqnarray}
  g^0  = \int d E \left(-{\partial f \over \partial E}\right)
  \left\{ x {\Gamma^2 \over (E-E_0)^2+\Gamma^2/4}
  + y \cos (2 \pi {\Phi \over \Phi_0} + \xi_0 ) + \right. \nonumber \\
   \left. z \sin \delta (E)
   \sin ( 2 \pi {\Phi \over \Phi_0} + \xi_R - \delta (E) ) \right\} ,
\label{E4}
\end{eqnarray}
with positive coefficients $x$, $y$, $z$ and real phase shifts $\xi_0$,
$\xi_R$. The resonance phase $\delta (E)$ is defined by
\begin{equation}
  \exp (i \delta) = - {E-E_0-i\Gamma/2 \over |E-E_0-i\Gamma/2 |}
\label{E5}
\end{equation}
and takes the value $\pi/2$ at $E=E_0$ while approaching $0$ for $E \rightarrow
-\infty$ and $\pi$ for $E \rightarrow \infty$, respectively. Averaging over
disorder can be done by averaging the phase shifts $\xi_0$ and $\xi_R$
over some suitable interval. With a suitable redefinition of the quantities
$y$, $z$, $\xi_o$ and $\xi_R$, this procedure leaves the {\it form} of Eq.\
(\ref{E4}) unchanged.

In the experimentally relevant regime $\Gamma \ll k_BT$, the energy
average over the Breit--Wigner term yields a thermally broadened
resonance $\propto \cosh^{-2} ((E_F-E_0)/2k_BT)$. The average over
the last term in Eq.\ (\ref{E4}) can easily be
done numerically. The result has the form $z (\Gamma / k_BT) B(E_F - E_0)
\sin (2 \pi \Phi / \Phi_0 + \xi_R-\beta(E_F - E_0))$. Both the amplitude
$B(E_F - E_0)$ and the phase $\beta(E_F - E_0)$ are plotted in Fig.\ 1
versus the argument $(E_F - E_0)/k_B T$. Both quantities vary on a scale
defined by the thermal energy $k_BT$. The amplitude $B$ attains its maximum
value at $E_F=E_0$ and vanishes with a power--law dependence for
$|E_F-E_0| \gg k_BT$. Comparing this with the exponential decay of the
temperature--averaged Breit--Wigner resonance term, we see that the
dominance of the interference term outside the resonance persists
after temperature--averaging. At $E_F=E_0$ the phase $\beta$ goes
through $\pi/2$. We note that in the {\it absence} of non--resonant
transmission,
i.e. for $t_{ab}^{L0} = 0$, it is the phase $\beta$ which would be measured
as the phase $\alpha$ in the AB experiment. Clearly, the rise of $\beta$
in the resonance region is not by far as steep as is observed experimentally
(compare Fig.\ 2 for the experimental data).
Collecting our results, we find for $g^0$
\begin{eqnarray}
  g^0 = x {\pi \Gamma \over 2 k_BT} \cosh^{-2} \left( {E_0-E_F \over
  2k_BT} \right)
  + y \cos (2 \pi {\Phi \over \Phi_0} + \xi_0 ) \nonumber \\
  + z {\Gamma \over k_BT} B(E_F - E_0) \sin \left(2 \pi {\Phi \over \Phi_0} +
    \xi_R-\beta(E_F - E_0)\right).
\label{E6}
\end{eqnarray}
{}From Eq.\ (\ref{E6}), it is immediately obvious that the measured phase
$\alpha$ must change by $2 \pi$ as the energy sweeps through a resonance, in
keeping with the first experimental finding. Indeed, far below and far above
the resonance, the first and third terms of Eq.\ (\ref{E6}) do not contribute
to the current, and the AB phase is there determined by the
energy--independent second term alone.
To determine the energy dependence of the measured phase $\alpha$ explicitly,
we write the sum of the two $\Phi$-dependent terms in Eq.\ (\ref{E6})
in the form $C(E_F - E_0) \cos (2 \pi \Phi / \Phi_0 + \xi_0 +
\alpha(E_F - E_0))$.
{}From Eq.\ (\ref{E6}) we find
\begin{equation}
  \tan \alpha(E_F - E_0) = -{z (\Gamma / k_BT) B(E_F - E_0)
  \cos \left[\xi_R-\xi_0-\beta(E_F - E_0)\right] \over
  y+z (\Gamma / k_BT)B(E_F - E_0) \sin
  \left[\xi_R -\xi_0-\beta(E_F - E_0)\right]}
\label{E7}
\end{equation}
To discuss the energy dependence of $\alpha$ qualitatively, we focus
attention on the denominator in Eq.\ (\ref{E7}). From our previous
discussion, we expect that $z \gg y$. On the other hand, we also have
$k_B T \gg \Gamma$, so that $z \Gamma / (k_B T y)$ may be of order unity:
{\it Averaging over temperature reduces the interference
term involving the resonance amplitude to a scale which may be comparable to
the interference term involving the non--resonant transmission amplitude}. In
this case, the denominator may vanish at or near the resonance peak (this
depends on the background phase $\xi_R - \xi_0$). Then, $\alpha$ becomes a
steep function of energy. We demonstrate this possibility in Fig.\ 2
which shows $\alpha(E_F - E_0)$ calculated for the following choice
of parameters. We use $\Gamma/k_BT =0.025$ as determined from experiment,
a small non--resonant transmission amplitude with $y/z = 0.005$, and
$\xi_R=\xi_0 =-\pi /2$. For comparison, we also show experimental data for
two different conductance resonances (crosses and hexagons), and the phase
$\beta(E_F - E_0)$ (dashed line) which would be measured in the absence of
non--resonant transmission through the left arm. The insert shows the phase
$\alpha$ over a larger interval. Our result reproduces the observed
sharp increase on a scale much smaller than $k_BT$. We emphasize that
the behavior of $\alpha$ as shown in Fig.\ 2 rests on the assumption
that $\xi_0 \approx \xi_R$. It changes qualitatively when the difference
between the two background phases is close to $\pi$.

The previous discussion rests on the assumption that the Coulomb
interaction between electrons is negligible beyond the Coulomb blockade
effect. We have generalized our model by including the interaction
between electrons in the dot in the mean--field approximation. The
structure of the formulas given above remains unchanged.

In summary, we have shown that the observed behavior of the phase $\alpha$
of the transmission amplitude through a quantum dot can be explained within
the model of independent electrons. Our explanation depends crucially
on the existence of a small but non--vanishing non--resonant transmission
amplitude through the dot. Our model can be tested experimentally, for
instance by a measurement of the phase and magnitude of the conductance
between resonances. Should such a test fail, we would conclude that
the experimental result can definitely not be accounted for in the framework
of non--interacting electrons. In this case, a many--body theory
including quantum fluctuations would be required.

{\bf Acknowledgements} We thank the authors of Ref. \cite{Ya94} for a
preprint of their work. One of us (H.\ A.\ W.) is grateful to Y. Gefen who
introduced him to the problem.

\begin{figure}

\begin{description}

\item[Fig.\ 1] The amplitude $B(E_F - E_0)$ (solid line) and the phase
$\beta(E_F - E_0)$ (dashed line) of the interference term involving the
resonance amplitude, and the transmission amplitude of the right arm,
both as functions of $(E_F-E_0)/k_BT$.

\item[Fig.\ 2] The energy dependence of the phase $\alpha(E_F - E_0) + \xi_0$
of the total transmission amplitude (solid line) calculated from Eq.\
(\ref{E7}) with $\Gamma/k_BT=0.025$, $y/z=0.005$ and $\xi_R = \xi_0= -\pi/2$.
For comparison,
the experimentally measured phases for two different resonances are shown
(crosses and hexagons). The dashed line shows the behavior of
$\alpha(E_F - E_0)$ for $y=0$, i.e. in the absence of non--resonant
transmission through the dot. The inset displays $\alpha(E_F - E_0)$
over a larger interval and shows the increase by $2 \pi$.

\end{description}

\end{figure}


\begin{references}

\bibitem{Ya94} A.\ Yacoby, M.\ Heiblum, D.\ Mahalu, and H.\ Shtrikman
(unpublished).

\bibitem{Lan} R.\ Landauer, IBM J.\ Res.\ Dev.\ {\bf 1}, 233 (1957);
Phil.\ Mag.\ {\bf 21}, 863 (1970); M.\ B\"uttiker, Phys.\ Rev.\ Lett.\
{\bf 57}, 1761 (1986).

\bibitem{AL} For a review, see D.\ V.\ Averin and K.\ K.\ Likharev, in
{\em Mesoscopic Phenomena in Solids}, edited by B.\ L.\ Altshuler,
P.\ A.\ Lee, and R.\ A.\ Webb, (Elsevier, Amsterdam, 1990).

\bibitem{B} For a review, see H.\ van Houten, C.\ W.\ J.\ Beenakker,
and A.\ A.\ M.\ Staring, in {\em Single Charge Tunneling}, edited
by H.\ Grabert and M.\ H.\ Devoret, (Plenum, New York, 1992).

\end{references}
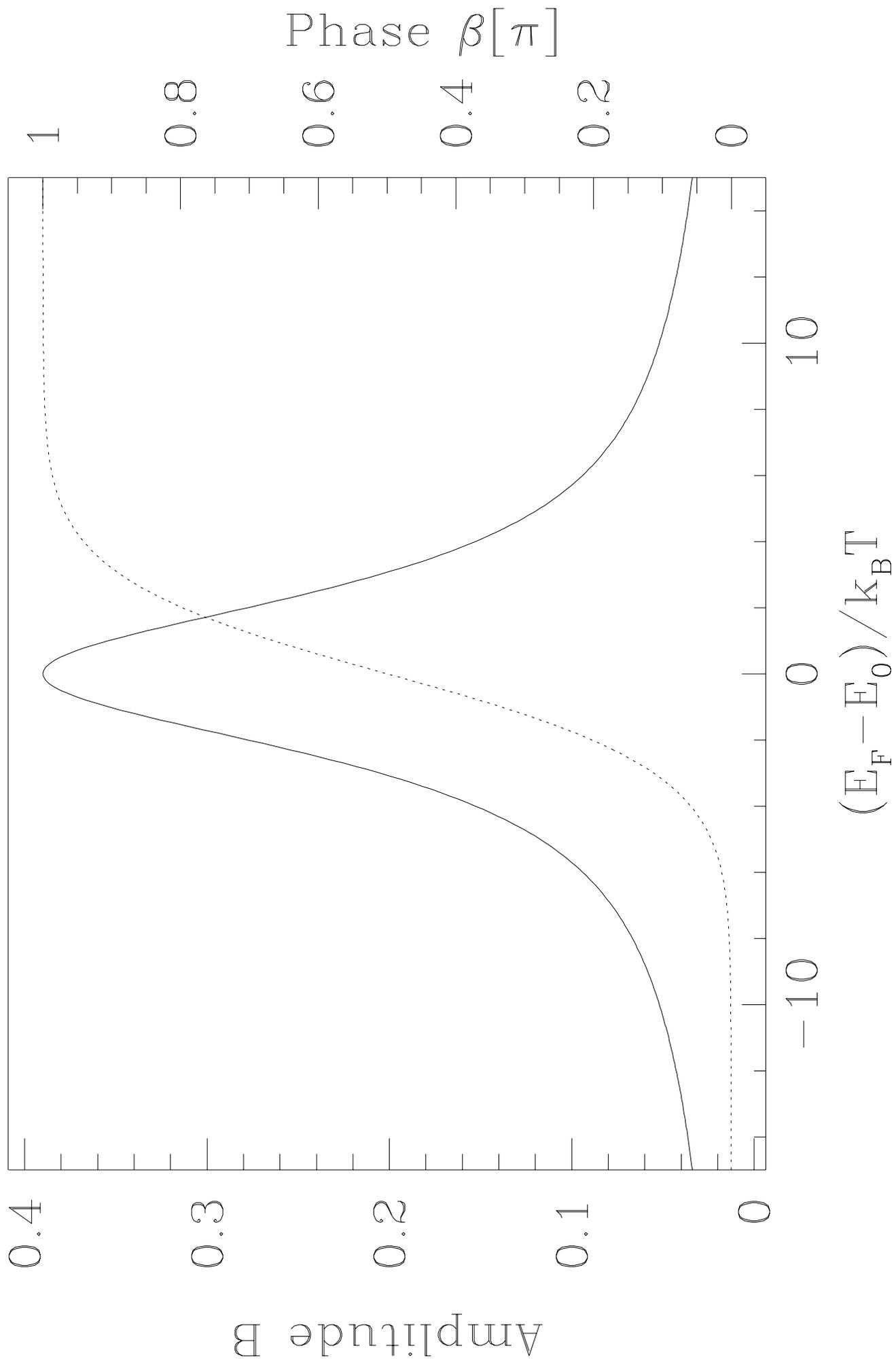
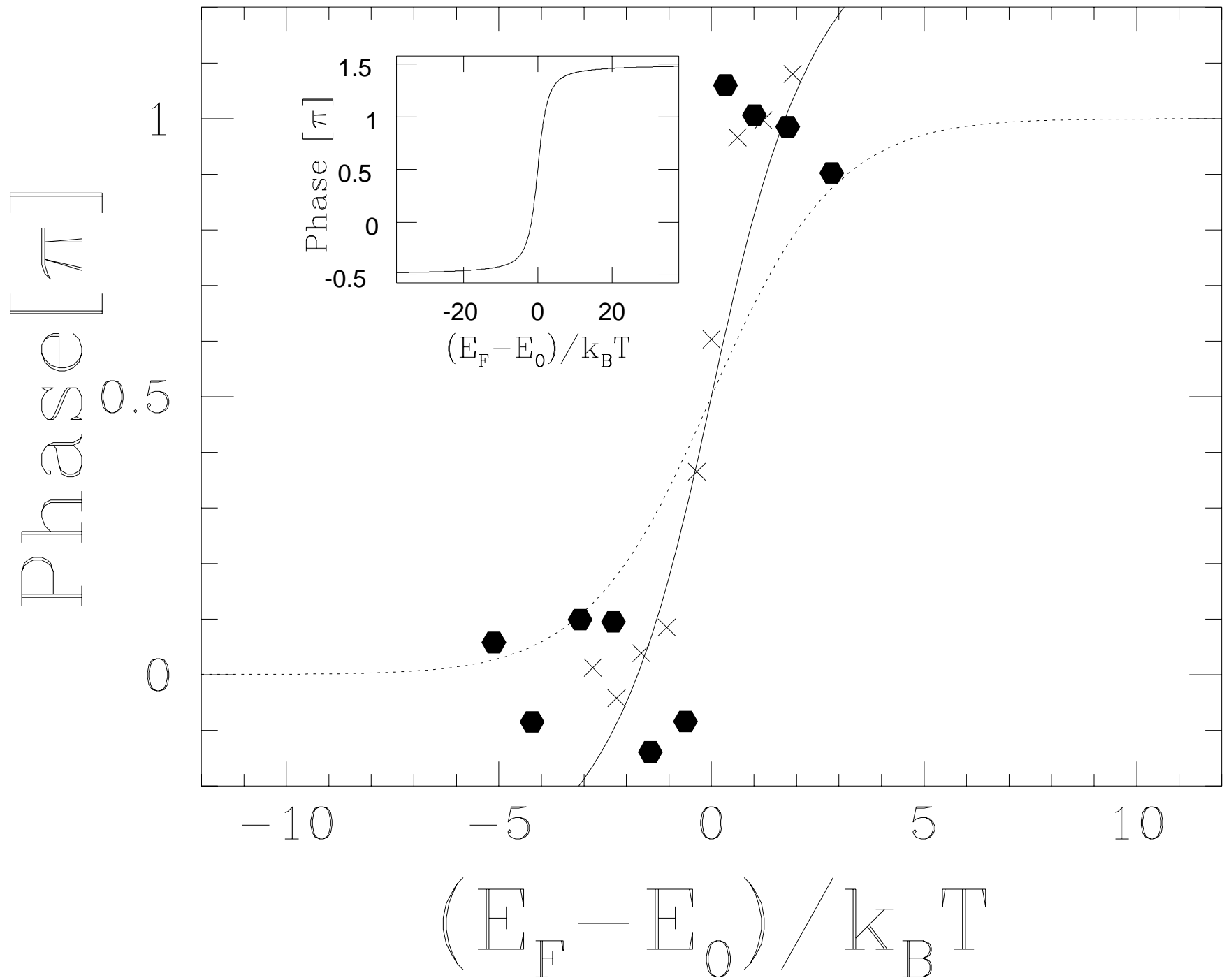
\end{document}